
\input phyzzx
\tolerance=10000

\def\papersize{\hsize=475pt \vsize=630pt \pagebottomfiller=0pt}
\papersize

\def\QQ{Q^2}
\def\mm{\mu^2}

\def\xii{\xi^2}

\def\al{\alpha(\mu^2)}

\def\DY{\scriptscriptstyle{\rm{DY}}}
\def\DIS{\scriptscriptstyle{\rm{DIS}}}
\def\MSb{\scriptscriptstyle{\overline{\rm{MS}}}}

\def\qqb{q\bar{q}}

\def\ep{\epsilon}
\def\eps{\epsilon}

\def\msb{\overline{\rm MS}}
\def\dis{{\rm DIS}}

\Pubnum = {ANL-HEP-25\cr
           CERN-TH-96-75\cr
           ITP-SB-96-17}

\titlepage
\title {SUDAKOV FACTORIZATION AND RESUMMATION}
\author{Harry Contopanagos$^1$,  
Eric Laenen$^2$ and George Sterman$^3$} 

\address{$^1$High Energy Physics
Division, Argonne National Laboratory\break
Argonne, IL 60439-4815}
\address{$^2$CERN Theory Division\break
CH-1211 Geneva 23, Switzerland}
\address{$^3$Institute for Theoretical Physics, SUNY Stony Brook\break
Stony Brook, New York 11794-3840}

\abstract
{We present a unified derivation of the resummation of  
Sudakov logarithms, directly
from the factorization properties of cross sections
in which they occur.  
We rederive in this manner the well-known exponentiation of
leading and nonleading logarithmic enhancements near the edge
of phase space for cross sections such as deeply inelastic 
scattering, which are induced by an electroweak hard scattering.
For QCD hard-scattering processes, such as heavy-quark production,
we show that the resummation of nonleading logarithms requires in general
mixing in the space of the color tensors of the hard
scattering.  The exponentiation of Sudakov logarithms implies 
that many weighted cross sections obey particular
evolution equations in momentum transfer, which streamline the
computation of their Sudakov exponents.
We illustrate this
method with the resummation of soft-gluon enhancements of the
inclusive Drell-Yan cross section, in both DIS and 
$\overline{{\rm MS}}$ factorization schemes.}

\vfill
\leftline{ANL-HEP-25}
\vskip -3pt
\leftline{CERN-TH-96-75}
\vskip -3pt
\leftline{ITP-SB-96-17}

\endpage
\pagenumber=1

\chapter{Introduction}

Our aim in this paper is to treat 
cross sections and amplitudes that allow Sudakov 
exponentiation and resummation
 in a 
unified and streamlined manner.  We shall point
out the factorization that underlies these
resummations, and derive the resummed form 
directly from it.  The close relation of 
factorization to Sudakov exponentiation was
emphasized and exploited long ago by
Mueller,
\Ref\Mueller{A.H.\ Mueller, Phys.\ Rev.\ D20 (1979) 2037.}\ 
Collins and Soper \Ref\CollinsSoper{J.C.
Collins and D.E.\ Soper, Nucl.\ Phys.\ B193 (1981) 381.}
and Sen.\Ref\Sen{A.\ Sen, Phys.\ Rev.\ D24 (1981) 3281.}  Here, we shall
generalize this observation,
and develop a method that makes the wide
range of its application to 
cross sections
and amplitudes straightforward and, we hope, clear.

A classic example\Ref\earlySud{V.V\
Sudakov, Sov.\ Phys.\ JETP 3 (1956) 65; M.\ Cassandro and M.\ Cini, Nuovo 
Cimento 34 (1964) 1719; R.\ Jackiw, Ann.\ Phys.\ (N.Y.) 48 (1968) 292;
P.M.\ Fishbane and J.D.\ Sullivan, Phys.\ Rev.\ D4 (1971) 458.} 
of Sudakov exponentiation 
is the dimensionally regulated electromagnetic form factor of
a massless quark\Ref\Collinsrv{J.C.\ Collins,
in {\it Perturbative Quantum Chromodynamics}, ed.\ A.H.\ Mueller (World Scientific,
Singapore, 1989).}
$$
\Gamma_\mu={\bar u}(p')\; \gamma_\mu\; u(p)\; F(Q^2,\epsilon)\, ,
\eqn\EMff
$$
with $Q^2=(p'-p)^2$.
  In QCD, the form factor
$F(Q^2,\epsilon)$ exponentiates as\Ref\MagSt{L.\ Magnea and G.\ Sterman,
Phys.\ Rev.\ D42 (1990) 4222.}
$$
F(Q^2,\epsilon)=\exp\; \Bigg \{ - {1\over 2}\; 
\int_0^{Q^2}{d\eta^2 \over \eta^2}\,
\Big  [ K(\alpha_s(\mu^2),\epsilon)
+
G\big ( \eta^2/\mu^2,\alpha_s(\mu^2),\epsilon \big ) 
\Big ]\; \Bigg \}\, ,
\eqn\FQsquar
$$
where, because $F(Q^2,\epsilon)$ is 
invariant under renormalization, the functions $K$ and $G$
satisfy
$$
\mu{dK\over d\mu}=-\mu{dG\over d\mu}\, .
\eqn\mdmKmdmG
$$
Infrared singularities are
summarized by  $K$, which contains at most a single pole in $\epsilon=2-d/2$,
while $G$, which summarizes ultraviolet
behavior, has at most a single logarithm in $\eta^2/\mu^2$.
$K$ and $G$ are each power series in $\alpha_s$, which
may be read off from low-order calculations. \refmark\MagSt\ \
For a fixed coupling, $\ln F(Q^2)$ has double logarithms of
$Q^2$, \refmark\earlySud and double poles in $\epsilon$,
 generated from the explicit integrals.

Sudakov exponentiation is much more general than this example,
however.  It applies as well to many cross sections 
involving a large momentum transfer,
in special limits of their phase spaces. As we shall
see, logarithms in these cross sections exponentiate 
after a Mellin or
Fourier transform.
We shall refer to this range of possibilities 
collectively as Sudakov
resummation.

The connection between factorization and resummation is 
already illustrated by 
perturbative renormalization, in
which the general relation of unrenormalized and
renormalized Green functions of fields $\phi_i$
carrying momenta $p_i$ is
$$
G_{\rm un}(p_i,M,g_0)=
\prod_i Z_i^{1/2} (\mu/M,g(\mu))\; G_{\rm ren}(p_i,\mu,g(\mu))\, .
\eqn\Gren
$$
$M$ is an  ultraviolet cutoff, and $g(\mu)$
and $g_0$ are the renormalized and bare couplings respectively.  
The independence of $G_{\rm un}$
from $\mu$ and $G_{\rm ren}$ from $M$ may be used to derive 
renormalization group equations, 
$$
\mu{d\ \ln\; G_{\rm ren}\over d\mu}=-\sum_i \gamma_i(g(\mu))\, ,
\eqn\rgG
$$
in which the anomalous dimensions $\gamma_i=(1/2)(\mu d/d\mu)\ln Z_i$
appear as constants in the
separation of variables, free of explicit dependence on
either $\mu$ or $M$. 

Very similar observations apply to structure functions in 
deeply inelastic scattering (DIS), which we may write in a 
convolution form, neglecting particle labels, as
$$
F(x,Q^2)
=
\int_x^1 {d y\over y}\, C(x/y,Q^2/\mu^2,g(\mu))\; \phi(y,\mu^2)\, ,
\eqn\Ffact
$$
where $C$ is a short-distance coefficient function, and 
$\phi$ is an infrared-sensitive parton distribution.
The parameter $\mu$ here plays the role of a ``factorization"
scale, separating long from short distances.
The expression \Ffact\ factorizes 
into a product under moments $\int_0^1 dx x^{N-1}$,
$$
{\tilde F}(N,Q^2)={\tilde C}(N,Q^2/\mu^2,g(\mu))
\; {\tilde \phi}(N,g(\mu))\, .
\eqn\dismt
$$
Since $\tilde F$ is physical and thus
independent of  $\mu$, we again
find a simple renormalization group equation for $\tilde C$ 
and $\tilde \phi$
$$
\bigg ( \mu{d\over d\mu}-\gamma_N(g) \bigg ) \ln{\tilde C} 
 =0=\bigg ( \mu{d\over d\mu}+\gamma_N(g) \bigg ) \ln{\tilde \phi}\, ,
\eqn\rgCphi
$$
where the anomalous dimension depends on $N$ and $g$,
the only dimensionless parameters shared by the two functions 
on the right-hand side of \dismt.

We will show how Sudakov 
behavior can be derived from a generalization of the
single-scale, two-function factorization illustrated
by DIS above, to a 
two-scale,
multi-function factorization. \Ref\CSSrv{J.C.\ Collins, D.E.\ Soper and 
G.\ Sterman,
in {\it Perturbative Quantum Chromodynamics}, ed.\ A.H.\ Mueller (World Scientific,
Singapore, 1989).}\Ref\CSSfactpapers{J.C.\ Collins, D.E.\ Soper and 
G.\ Sterman, Nucl.\ Phys.\ B261 (1985) 104; Nucl.\ 
Phys.\ B308 (1988) 833.}\Ref\Bodwinfact{G.T.\ Bodwin, Phys.\ Rev.\ 
D31 (1985) 2616.}  Such 
a factorization is characteristic of physical quantities 
that describe hard scatterings, and which 
are sensitive to three
generic regions of momentum space, associated with 
three sorts of quanta.  These are:  off-shell 
(``hard") partons at short distances,
fast-moving on-shell (``collinear") partons near the light cone, 
and long-wavelength, soft
partons.  The separation of a cross section into separate
functions for each type of excitation requires
multiple factorizations. The independence
of the final expression from the details of these
factorizations, however, leads to
a consistency equation, and consequently to
Sudakov resummation.  This process is related to the use of
matching conditions in effective field theories.
\Ref\Effective{K.G.\ Wilson and J.\ Kogut,
Phys.\ Rep.\ 12 (1974) 75; 
      J.\ Polchinski,
Nucl.\ Phys.\ B231 (1984) 269; H.\ Georgi,
 Ann.\ Rev.\ Nucl.\ Part.\ Sci., 43 (1993) 209.}  

A number of phenomenologically useful and theoretically
important resummations of this type have been derived or postulated
for quantities of experimental and theoretical
interest, including the 
Sudakov form factor itself, \refmark\Collinsrv\refmark\Mueller
\refmark\Sen\
the transverse momentum 
distribution of back-to-back particles in 
${\rm e}^+{\rm e}^-$ annihilation \refmark\CollinsSoper\Ref\bbjresum{J.C.\ 
Collins and D.E.\ Soper,
 Nucl.\ Phys.\ B197 (1982) 446.} and of
Drell-Yan pairs, \Ref\qtresum{J.C.\ Collins, D.E.\ Soper and G.\ Sterman, 
Nucl.\ Phys.\ B250 (1985) 199;
 G.\ Altarelli, R.K.\ Ellis, M.\ Greco and G.\ Martinelli,
Nucl.\ Phys.\ B246 (1984) 12;
   C.T.H.\ Davies and W.J.\ Stirling, Nucl.\ Phys.\ B244 (1984) 337;
   C.T.H.\ Davies, B.R.\ Webber and W.J.\ Stirling, Nucl.\ Phys.\ B256
    (1985) 413;
   P.B.\ Arnold and R.P.\ Kauffman, Nucl. Phys. B349 (1991) 381;
   G.A.\ Ladinsky and C.P.\ Yuan, Phys.\ Rev.\ 50 (1994) R4239.}\ the total
Drell-Yan cross section 
\Ref\Stdy{G.\ Sterman, Nucl.\ Phys.\ B{281} (1987) 310.}\Ref\CT{S. Catani 
and L. Trentadue, Nucl.\ Phys. B327 (1989) 
323.}\Ref\Catcomm{S. Catani and
L. Trentadue, Nucl.\ Phys.\ B353 (1991) 183.}
\Ref\CS{H. Contopanagos and G. Sterman, Nucl.\ Phys.\ B400 (1993)
211; 419 (1994) 77; L. Alvero and H. Contopanagos, Nucl.\ Phys.\ B436 (1995) 184.}
\Ref\AC{L. Alvero and H. Contopanagos, Nucl.\ Phys.\ 
B456 (1995) 497.}
and a large class of infrared safe event
shapes in ${\rm e}^+{\rm e}^-$ annihilation,
\Ref\CTTW{S.\ Catani, L.\ Trentadue, G.\ Turnock and B.R.\ Webber, 
     Nucl. Phys. B407 (1993) 3.}
 as well
as elastic scattering amplitudes at very high energy.
\Ref\BSt{J.\ Botts and G.\ Sterman, Nucl.\ Phys.\ B325 (1989) 62.}\
Closely related expressions
apply to the momentum dependence of vacuum expectations of
products of Wilson lines (ordered exponentials)
\Ref\KRad{G.P.\ Korchemsky and A.V.\ Radyushkin, Phys.\ Lett.\ B171 (1986) 459;
I.A.\ Korchemskaya and G.P.\ Korchemsky,
Nucl.\ Phys.\ B437 (1995) 127; J.G.M.\ Gatheral, Phys.\ Lett.\ B113 (1983) 90;
J.\ Frenkel and J.C.\ Taylor, Nucl.\ Phys.\ B246 (1984) 231.}.
Although all related, the derivations
of Sudakov resummations for these quantities 
have been
sufficiently varied in presentation and emphasis that
a single derivation from an underlying principle
should be useful.    
In this paper, we shall
provide such a derivation based on a common
factorization into 
functions sensitive to hard, collinear and soft partons.

Yet another motivation for studying resummations of this
form is that they naturally result in exponentials of integrals over
momentum scales of the running coupling.  The 
singularity of the perturbative $\alpha_s(\mu^2)$ at 
$\mu=\Lambda_{\rm QCD}$ is sometimes referred to as
an infrared renormalon 
\Ref\IRR{D.\ Appell, P.\ Mackenzie and 
G.\ Sterman in {\it 
Proceedings of the Storrs Meeting}, Fourth Meeting of the Division
of Particles and Fields, Storrs, CT, August 15-18, 1988, p.\ 567;
 B.R.\ Webber, Phys.\ Lett.\ B339 (1994) 148;  
G.P.\ Korchemsky and G.\ Sterman, Phys.\ Lett.\ B340 (1994) 96;
Nucl.\ Phys.\ {B437} (1995) 415;
A.V.\ Manohar and M.B.\ Wise, Phys.\ Lett.\ {B344} (1995) 407;
Yu.L.\ Dokshitzer and B.R.\ Webber, 
        Phys.\ Lett.\ {B352} (1995) 451;
   R.\ Akhoury and V.I.\ Zakharov, Phys.\ Lett.\ {B357} (1995) 646;
 P.\ Nason and M.H.\ Seymour, Nucl.\ Phys.\ {B454} (1995) 291;
M.\ Beneke and V.M.\ Braun, Nucl.\ Phys.\ {B454} (1995) 253.}
, whose presence and
nature may shed light on power-suppressed  corrections
to cross sections of interest.  Although we shall not
investigate renormalons directly in this paper, the
generality of Sudakov resummation should help
emphasize their relevance.

The factorization
into short-distance, jet and soft functions
is quite general,
 applicable to many weighted
cross sections with large momentum transfers,
\refmark\CSSrv
and to some elastic amplitudes as well,
when considered in impact parameter form. \refmark\BSt\ \ 
In the following section
we shall use this
factorization to derive the general Sudakov resummation
formula for electroweak-induced
processes (Drell-Yan, deeply inelastic
scattering, ${\rm e}^+{\rm e}^-$).
This is the central result of this paper, from which we draw applications
in Sections 3-5.
Section 3 briefly discusses the extension of the formalism
to QCD hard scattering \refmark\BSt\Ref\KidSt{N.\ Kidonakis and
G.\ Sterman, Stony Brook preprint ITP-SB-96-7 (April, 1996),
hep-ph/9604234.}, which shows that
exponentiation applies as well to cross sections involving
jets and heavy quark production.
In Section 4 we discuss the reexpression of our 
result as an evolution equation
in terms of momentum transfer, whose solution is useful for the computation of
dimensionally-regularized cross sections and amplitudes.
Finally, in Section 5 we use the Drell-Yan and
deeply inelastic scattering cross sections as
examples.

\chapter{Resummation from Factorization}

This section begins with a review of the basic
properties held in common by all the 
cross sections (and, with slight modifications, amplitudes)
to which Sudakov resummation applies.
 These properties are summarized in a rather specific but relatively simple
 factorized expression satisfied by each cross section in a particular
limit of its final-state phase space
\refmark\CSSrv.  This underlying factorization is
illustrated in Fig\ 1 for ${\rm e}^+{\rm e}^-$ annihilation
(a), the Drell-Yan cross section (b) and DIS (c).  The figure is
to be thought of as a reduced diagram, showing on-shell
lines of the amplitude and its complex conjugate, in cut diagram notation. 
All lines are massless.

In the figure, electroweak vector bosons
of large spacelike or timelike mometum $q$, $q^2=\pm Q^2$,
are linked to partons through two
hard-scattering functions $H$ and $H^*$, in 
the amplitude and the complex conjugate respectively.
$H$ and $H^*$ depend only on quanta off-shell by order $Q^2$. 
On-shell particles with momenta of order $Q$
fall into two jets, $J_1$ and $J_2$ of
collinear particles.  We assume that
the final state, represented by the cut of the diagram, approaches an
``edge" of phase space, where all finite-energy particles
are concentrated within the two jets.  We shall also  refer to this as an
``elastic" limit, although the final
states include arbitrary numbers of particles.
In this terminology, the diagrams of Fig.\ 1 represent the most general 
configurations that contribute at leading power in $Q^2$
 in the elastic
limit. \refmark\CSSrv

As shown in the figure, jets may be produced at
the hard scattering, as in ${\rm e}^+{\rm e}^-$ 
annihilation (Fig.\ 1a), or they may originate with an 
incoming parton, 
and hence be identified with a parton
distribution, as in the Drell-Yan process (Fig.\ 1b).
In deeply inelastic scattering (Fig.\ 1c), we have one
parton distribution and one outgoing jet.
In addition, each such  cross section involves
the emission of soft particles, represented
by function $S$.  
The elastic limit may be isolated quantitatively by
introducing an appropriate weight.

By a weight, we refer to a set of functions $w_n(\{k_i\})$ of 
the momenta of the particles in 
each $n$-particle final state, $\{k_i\}$, $i=1\dots n$.
The weighted cross section is a sum over all
final states, each integrated over its
phase space with weight $w_n(\{k_i\})$.  
For definiteness,
we assume that the $w_n$ are dimensionless, and that they
vanish in the elastic limit identified  above.
We also assume the choice of weight
is consistent with the cancellation of infrared divergences
due to final-state interactions.  We shall refer to such a weight 
as being
{\it infrared safe}.  This requirement is met if the $w_n(\{k_i\})$ are
symmetric in the momenta $k_i$ and satisfy the
relations \Ref\IRSweight{G.\ Tiktopoulos, Nucl.\ Phys.\ B147 (1979) 371; 
G.\ Sterman, Phys.\ Rev.\ D19 (1979), 3135.}
$$
w_n(k_1,\dots  (1-\alpha)k_{n-1},\alpha k_{n-1})
=
w_{n-1}(k_1,\dots k_{n-1})\, ,
\eqn\wirs
$$
which ensures that two states that differ
by the emission of zero-momentum particles,
or by the rearrangement of momenta
among collinear particles, have the same weight. 

In $n$th-order perturbation theory, we encounter
 logarithmic enhancements in the elastic limit
on a diagram-by-diagram basis, with
$\alpha_s^n\; [{(\ln^{2n-1}w)\; /\; w} ]$
typically the most singular behavior at order $\alpha_s^n$.
These are the singularities we shall organize.
In dimensionally-regularized perturbation theory,
these logarithms appear from the expansion of
kinematic combinations such as 
$$
{1\over\epsilon}{1\over {\big [ wQ^a\big ]}^{1+\epsilon}}\, ,
\eqn\eQcombo
$$
with $a=1$ or 2 (and $\epsilon=2-d/2$), examples of which we shall encounter below.

Not every infrared safe weight leads to exponentiation, however \refmark\CTTW.
We will need to assume in addition that near the elastic limit,
the contributions to the weight of particles within the jets and
soft function are independent and additive,
$$
w=w_1+w_2+w_s\, ,
\eqn\weightadd
$$
with corrections that vanish as $w^2$ for
small $w$.  

Consider the processes of Fig\ 1.
For the ${\rm e}^+{\rm e}^-$ annihilation cross section,
we may identify $w=1-T$, with $T$ the thrust \refmark\CTTW,
although similar enhancements may occur in 
other variables as well.
For the inclusive
Drell-Yan cross section \refmark\Stdy, 
enhancements occur in the variable 
$w=1-Q^2/s\equiv 1-\tau$, 
while for incusive DIS they occur
in $1-x$, with $x$ the Bjorken scaling variable \refmark\Stdy.
  Finally, for the Drell-Yan cross section at measured
transverse momentum \refmark\qtresum
$d\sigma/dQ^2d^2Q_T$, we may take $w\sim Q_T/Q$. 

In the elastic limit, we associate with the jets
 momenta $p_1$ and $p_2$, with $Q^2=\pm(p_1\pm p_2)^2$
the scale of the hard scattering.
For two outgoing jets, as in ${\rm e}^+{\rm e}^-$ annihilation,
we may regard each $p_i^\mu$ as the lightlike jet
momentum in the elastic limit, and $Q^2=(p_1+p_2)^2$.  
For an incoming
jet, as in DIS, $p_1$
may be taken as the lightlike
momentum of the initial-state parton.  In this case,
$Q^2=-q^2=-(p_1-p_2)^2$.

The general factorization form near
$w=0$, applicable to all these cross sections is
$$
\eqalign{
\sigma(w)
=&
H(p_1/\mu,p_2/\mu,\zeta_i)
\int {dw_1\over w_1}\; {dw_2\over w_2}\; {dw_s\over w_s}\cr
 &\hbox{\hskip 1.5 true in} \times
 J_1(p_1\cdot \zeta_1/\mu,w_1(Q/\mu)^{a_1}) 
J_2(p_2\cdot \zeta_2/\mu,w_2(Q/\mu)^{a_2})
\cr
 &\hbox{\hskip 1.5 true in} \times
S(w_s(Q/\mu),v_i,\zeta_i)\; \delta(w-w_1-w_2-w_s)\, ,
}
\eqn\Sudconv
$$
with corrections that vanish as powers of $w$.
The specific kinematics in \Sudconv\ is defined by
$$
p_1=Q^+v_1,\ p_2=Q^-v_2,\ q^2=\pm Q^2=\pm 2Q^+Q^-\, ,
\eqn\kinemat
$$
with $v_i$ dimensionless lightlike vectors.
As indicated, we shall take $v_1^\mu=\delta_{\mu +}$,
$v_2^\mu=\delta_{\mu -}$ for definiteness.  
The factorization eq.\ \Sudconv\ often applies even if the cross section
is not infrared safe, but may be factorized according
to the standard procedure. \refmark\CSSrv  In Drell-Yan, for instance,
collinear divergences  are factored as usual into the
jets $J_i$, which play the role of parton distributions, as
noted above.
In \Sudconv, $J_i$, $S$ and $H$ represent
the contributions of the two jets, of the soft quanta
not part of the jets, and of the hard quanta, respectively.
We have explicitly exhibited overall factors of $1/w_i$, $1/w_s$,
which we readily derive from power counting for 
the soft and jet functions in \Sudconv.\refmark\Stdy

We may think
of the vectors $\zeta_1$ and $\zeta_2$
as gauge-fixing vectors $\zeta_i\cdot A=0$, used to
define the jets $J_i$ in terms of particular 
 matrix elements. \refmark\Stdy
In such axial gauges, the 
factorization of Sudakov double logarithms into the jets is 
automatic \refmark\Collinsrv.  
Although the jet
definitions may be gauge-dependent, the cross
section itself, of course, remains gauge invariant.

The arguments $w_i(Q/\mu)^{a_i}$
represent the combination of weight and momentum
transfer $Q$ through which the jets depend upon the
weight.  The value of $a_i$ in a given cross section depends on
the underlying kinematics.  In the analysis of the
$x\rightarrow 1$ limit of DIS in Ref.\ [\Stdy] (where $w=1-x$),
for instance, $a=1$ for the initial-state jet  and $a=2$ for the 
final-state jet.
We assume (as is
the case for all the examples we know of) that the soft
function depends only on the variable
$w_s Q$.  In principle, other combinations of $Q$ and 
 $w$-dependence in eq.\ \Sudconv\ are possible, but we shall not describe
them here.
Because of their $\zeta_i$-dependence, the  $J_i$ depend on 
$p_i\cdot\zeta_i$ and, in general, $\zeta^2$. 
The variables $p_i\cdot\zeta_i$, which
depend on $\zeta_i$, are  independent of $w_iQ^{a_i}$, which is defined by
kinematics.

The convolution \Sudconv\ factorizes into a simple product under
the transformation 
$$
\eqalign{
{\tilde \sigma}(N) =
\int_0^\infty dw\, e^{-Nw} &=H(p_1/\mu,p_2/\mu,\zeta_i)\, {\tilde S}(Q/\mu N,v_i,\zeta_i) 
\sigma(w) \cr
& \quad\quad \times
{\tilde J}_1(p_1\cdot \zeta_1/\mu,Q/\mu N^{1/{a_1}})\;  
{\tilde J}_2(p_2\cdot \zeta_2/\mu,Q/\mu N^{1/{a_2}})\, ,
}
\eqn\transform
$$
where, for instance, ${\tilde J}=\int_0^\infty\; (dw/w)\; \exp[-Nw]\; J$.
Singular behavior in
the $w\rightarrow 0$ limit is reflected by
growth for large values of the conjugate variable $N$.
This is the significance of the lower limit on the
$w$ integral.  For large $N$, contributions from
the upper limit are suppressed, and it 
may be extended  to infinity.
We consider $N$ as a complex variable, so that
\transform\ represents both Laplace and Fourier transforms.
For large $|N|$, it serves as well as a representation
of the Mellin transform, since 
for $w\rightarrow 0$, $\exp[-Nw]\sim (1-w)^N$, with corrections
that vanish as $w$, which is the accuracy of the
basic factorization formula \Sudconv.

In general, 
the individual functions $H$, $J_i$ and $S$ each
require renormalization, which we assume to be
multiplicative.  
Their renormalization scale dependence is 
determined by a set of anomalous dimensions,
$$
\eqalign{
\mu {d \over d\mu} \ln H &= -\gamma_H(\alpha_s)\ \cr
\mu {d \over d\mu} \ln J_i &= -\gamma_{J_i}(\alpha_s) \cr
\mu {d \over d\mu} \ln S &= -\gamma_S(\alpha_s)\, .
}
\eqn\anomdim
$$
We assume for 
this discussion that these
anomalous dimensions are independent of the moment variable $N$
and of the $\zeta$'s.
The independence of $\tilde{\sigma}(N)$ from
the renormalization scale $\mu$ leads to 
$$
\eqalign{
\gamma_H+\gamma_S+\sum_i \gamma_{J_i} &=0\, ,
}
\eqn\sumanomequalzero
$$
that is, the renormalization-dependence of the individual
functions must cancel in their product.

In summary, we are concerned here with weighted cross sections that
satisfy the short distance--light cone--long distance convolution
of eq.\ \Sudconv\ in terms of functions that are renormalizable
according to eq.\ \anomdim. 
It seems likely
that \Sudconv\ has an interpretation in the language of effective 
field  theories, \refmark\Effective\ applied 
to massless particles in Minkowski space.

We emphasize that in this paper we shall not prove eq.\ 
\Sudconv\ and its related renormalization conditions \anomdim\
for any process, nor do we wish to minimize the
challenge involved in carrying out such a proof in any particular 
case.  
For the status of such proofs, we refer the reader to the 
literature cited above.
We shall simply assume the result, and show that it
is restrictive enough to imply very specific resummation
formulas for the small-$w$ limit.

In the arguments below, we shall treat the choices of
$\zeta_i$ as analogous to an
ambiguous choice of renormalization
scheme.  Like a choice of factorization scale, this
very ambiguity results in strong restrictions on 
the cross section.

The general factorization eq.\ \Sudconv, then, cannot depend on
the precise choice of $\zeta_1$ or $\zeta_2$. 
A small
change of $\zeta_1$, for instance,
leads to a relation between the
first-order changes in $H$, ${\tilde S}$ and 
${\tilde J}_1$, which must cancel. Thus, under a variation 
of $p_1\cdot\zeta_1$ at fixed $\zeta_1^2$ (and
any other implicit $\zeta_1$-dependence) in the functions, we have the relation
$$
0= \bigg (p_1\cdot\zeta_1
{\partial \over \partial p_1\cdot\zeta_1} H\;\bigg)
{\tilde J}_1\; 
{\tilde J}_2\; {\tilde S}
+ H\; \bigg ( p_1\cdot\zeta_1
{\partial \over \partial p_1\cdot\zeta_1} 
{\tilde J_1}
\bigg )\; 
{\tilde J}_2\; {\tilde S}
+ H\; {\tilde J}_1\; {\tilde J}_2\; 
\bigg ( p_1\cdot\zeta_1{\partial \over \partial p_1\cdot\zeta_1} 
{\tilde S} \bigg )\, ,
\eqn\zetachange
$$
(${\tilde J}_2$
is independent of $\zeta_1$) or equivalently, 
$$
\eqalign{
p_1\cdot\zeta_1{\partial \over \partial p_1\cdot\zeta_1}
 \ln {\tilde J}_1(p_1\cdot\zeta_1/\mu,Q/\mu N^{1/a_1})
&=
-p_1\cdot\zeta_1{\partial \over \partial p_1\cdot\zeta_1}
 \ln H(p_1/\mu,p_2/\mu,\zeta_i) \cr
& \quad \quad \quad 
-p_1\cdot\zeta_1{\partial \over \partial p_1\cdot\zeta_1}
 \ln {\tilde S}(Q/\mu N,v_i,\zeta_i)\, .
}
\eqn\matchchange
$$
The logarithmic derivative of ${\tilde J}_1$ with respect to
$p_1\cdot\zeta_1$ may depend on the hard scale
$Q\sim p_1\cdot\zeta_1$, or the soft scale $Q/N$.
From \matchchange, this dependence is
additive in terms of one function that depends on the common
variables of ${\tilde J}_1$ and $H$ and another
that depends on the common variables of ${\tilde J}_1$ and ${\tilde S}$.
For $H$ these are $\alpha_s(\mu^2)$ and $p_1\cdot\zeta_1/\mu$.
For ${\tilde S}$ they are $\alpha_s(\mu^2)$ and another variable that depends upon
$a_1$, given by
$$
{1 \over (v_i\cdot\zeta_i)^{a_1-1}}\; {Q\over\mu N}
={Q^{a_1}\over ({p_i\cdot\zeta_i})^{a_1-1}\mu N}
\equiv{Q'_{a_1} \over \mu N}\, ,
\eqn\qsubadef
$$
where the final form serves to define $Q'_{a_1}$.

We may thus write \matchchange\ as
$$
p_1\cdot\zeta_1{\partial \over \partial p_1\cdot\zeta_1} 
\ln J_1(p_1\cdot \zeta_1/\mu,Q/\mu N^{1/a_1})
=
G(p_1\cdot\zeta_1/\mu,\alpha_s(\mu^2))
+K(Q'_{a_1}/\mu N,\alpha_s(\mu^2))\, ,
\eqn\GplusK
$$
in which the function $G$ matches the variation of the hard
part and $K$ the variation of the soft part,
$$
\eqalign{
G = &
-p_1\cdot\zeta_1{\partial \over \partial p_1\cdot\zeta_1}\ln H\, , \cr
K = &
-p_1\cdot\zeta_1{\partial \over \partial p_1\cdot\zeta_1}\ln {\tilde S}\, .
}
\eqn\GKdef
$$
Eq.\  \GplusK\ is the basic consistency condition referred
to in the introduction.

Because $J$ is multiplicatively renormalized,
the combination $G+K$ is itself a renormalization
invariant \refmark\CollinsSoper\refmark\Sen
$$
p_1\cdot \zeta_1 {d\over d p_1\cdot \zeta_1}\; \gamma_{J_1}(\alpha_s)
=
\mu{d\over d\mu}
 \bigg [G \big (p_1\cdot\zeta_1/\mu,\alpha_s(\mu^2)\big )
+K \big (Q'_{a_1}/\mu N,\alpha_s(\mu^2) \big ) \bigg ]=0\, ,
\eqn\GKrg
$$
or, again separating variables, \refmark\CollinsSoper
$$
\eqalign{
\mu{d\over d\mu}
K \big (Q'_{a_1}/N\mu,\alpha_s(\mu^2)\big )=&-\gamma_K \big (\alpha_s(\mu^2) \big )\, ,
\cr
\mu{d\over d\mu}
G \big (p_1\cdot\zeta_1/\mu,\alpha_s(\mu^2) \big )
=&\ \gamma_K \big (\alpha_s(\mu^2) \big )\, ,
}
\eqn\gammaKdef
$$
with $\gamma_K$ a Sudakov anomalous dimension.
 This
relation allows us to reexpress $G+K$ in a form which relates the
scales $p_1\cdot \zeta_1$ and $Q'_{a_1}/N$,
$$
K(Q'_{a_1}/N\mu,\alpha_s(\mu^2)))
+
G(p_1\cdot\zeta_1/\mu,\alpha_s(\mu^2))
=
- \int_{Q'_{a_1}/N}^{p_1\cdot\zeta_1}
 {d\mu' \over \mu'}\, A(\alpha_s({\mu'}^2))
+ A'\big (\alpha_s((p_1\cdot \zeta_1)^2) \big )\, ,
\eqn\GKrgsoln
$$
where the functions $A$ and $A'$ are defined by
$$
A(\alpha_s)=\gamma_K(\alpha_s)
+\beta(g){\partial \over \partial g} K(1,\alpha_s)\, ,
\eqn\donedef
$$
and
$$
A'(\alpha_s)=K(1,\alpha_s)
+
G(1,\alpha_s)\, .
\eqn\dtwodef
$$

A solution to both \GplusK\ and the renormalization group equation
for the jets in 
\anomdim, which organizes both 
$p_i\cdot\zeta_i$-
and $N$-dependence is
$$
\eqalign
{
{\tilde J}(p\cdot\zeta/\mu,Q/\mu N^{1/a},&\alpha_s(\mu^2))
=
{\tilde J}(1,1,\alpha_s(Q^2/N^{2/a}))\, 
\exp\; \Big [ - \int_{Q/N^{1/a}}^\mu 
{d\lambda \over \lambda}\gamma_J(\alpha_s(\lambda^2))\Big ]
\cr
&\times
\exp\Big [ - \int_{Q/N^{1/a}}^{p\cdot\zeta}{d\lambda \over \lambda}
\Big\{\int_{Q^a/\lambda^{a-1}N}^\lambda {d\xi \over \xi}\; A(\alpha_s(\xi^2))
- A'(\alpha_s(\lambda^2)\Big\}\, \Big ]\, ,
}
\eqn\tildeJsoln
$$
where we have used the definition in \qsubadef\ for $Q_a'$.  

In the following, we take $p\cdot\zeta=Q$.  Shifts of this value for
the form $p\cdot\zeta=CQ$, with $C$ a constant, may be absorbed into
redefinitions of $A'$ \refmark\CollinsSoper.
Finally, using the renormalization group equations \anomdim\ for $H$ and $S$,
and the relation 
\sumanomequalzero\ between their anomalous dimensions, 
we find
$$
\eqalign{
\ln\, {\tilde \sigma}(N) 
&=D_1(\alpha_s(Q^2))+D_2(\alpha_s(Q^2/N^2),a) \cr
&\ \hbox{\hskip 1.0 true in}
- 2\int_{Q/N^{1/a}}^{Q}{d\lambda \over \lambda}
\bigg [
\int_{Q^a/\lambda^{a-1}N}^\lambda {d\xi \over \xi}\; A(\alpha_s(\xi^2))
- B(\alpha_s(\lambda^2)) \bigg ]
\cr
&=D_1(\alpha_s(Q^2))+D_2(\alpha_s(Q^2/N^2),a) \cr
&\ \hbox{\hskip 1.0 true in}
- 2\int_{Q/N^{1/a}}^{Q}{d\xi \over \xi}
\bigg [
\ln(Q/\xi)\, A(\alpha_s(\xi^2))
- B(\alpha_s(\xi^2)) \bigg ]\cr
&\ \hbox{\hskip 1.0 true in}
-{2\over a-1}\, \int_{Q/N}^{Q/N^{1/a}}{d\xi \over \xi}\, 
\ln(\xi N/Q)\, A(\alpha_s(\xi^2))\, ,
}
\eqn\logsigmasoln
$$
where for simplicity we have taken $a_1=a_2=a$, and where
$$
\eqalign{
D_1(\alpha_s(Q^2))
&=
\ln H(1,1,\alpha_s(Q^2)), \cr
D_2(\alpha_s(Q^2/N^2),a)
&=
\ln {\tilde S}(1,\alpha_s(Q^2/N^2))
+
\sum_{i=1,2}\ln {\tilde J}_i(1,1,\alpha_s(Q^2/N^{2/a}))\cr
&\ \hbox{\hskip 1.0 true in}
-\int_{Q/N}^{Q/(N)^{1/a}}{d\lambda\over \lambda}
\gamma_S(\alpha_s(\lambda^2)), \cr
B(\alpha_s)
&=
{1\over 2}\gamma_H+A'\, .
}
\eqn\sdef
$$
Note that the last term on the right-hand side of \logsigmasoln\ vanishes for
$a=1$, in spite of its overall factor of $1/(a-1)$.  Also, we note
that in cross sections involving partons in the initial state,
such as DIS and Drell-Yan, the function $D_2$ will in general
contain collinear singularities ($1/\epsilon$ in dimensional 
regularization).

At one loop in $\alpha_s$, denoting $f(\alpha_s)=
(\alpha_s/\pi)f^{(1)}+\dots$, we have
$$
\ln {\tilde \sigma}(N)
= D_1^{(0)}+D_2^{(0)}+(\alpha_s/\pi)
\bigg [ D_1^{(1)}+D_2^{(1)} - A^{(1)}\bigg({1\over a}\bigg)\ln^2N
+ 2 B^{(1)}\bigg({1\over a}\bigg )\ln N \bigg ]\, .
\eqn\oneloop
$$
Thus, from the one-loop cross section the functions $A$ and $B$
and the combination $D_1+D_2$ may be simply read off.
Higher orders in the functions are similarly determined to
higher loop order.

The result \logsigmasoln\ shows that the cross section is an exponential,
whose exponent has at most double logarithms of the weight
variable $N$ for fixed coupling, with higher powers of $N$
generated only by expansion of the running coupling.  This
is the basic 
Sudakov exponentiation, in the form derived, for
example, by Collins and Soper \refmark\CollinsSoper,
 and in Ref.\ [\BSt].
The $N$-dependence of couplings in the function $D_2$ is, of course,
consistent with these results.  These formulas, however,
must be restricted to values of $N$ for which $\alpha_s(Q^2/N^2)$
remains well-defined.  To go beyond this region, at least formally,
we shall use the solution to an evolution equation
in momentum transfer, which we shall derive in Section 4 below.
Before doing so, however, we observe how these results may be
generalized to purely QCD hard-scattering processes, such as inclusive
heavy-quark or jet production.

\chapter{Sudakov Resummation for QCD Hard Scattering}

The methods of 
Section 2 above apply rather directly to processes 
that proceed through color
exchange, such as heavy quark production
 \Ref\QQprod{E.\ Laenen, J.\ Smith and W.L.\ van Neerven, 
Nucl.\ Phys.\  B369 (1992) 543;
Phys.\ Lett.\ B321 (1994) 254;
E.L.\ Berger and H.\ Contopanagos, Phys.\ Lett.\  B361 (1995) 115;
 Argonne preprint ANL-HEP-95-82,
hep-ph/9603326;
N.\ Kidonakis and J.\ Smith, Phys.\ Rev.\ D51 (1995) 6092; 
S.\ Catani, M.L.\ Mangano, P.\ Nason and L.\ Trentadue,
CERN preprint CERN-TH/96-21, hep-ph/9602208.}\refmark\KidSt.  
To our knowledge, however, 
until recently \refmark\KidSt\ the only treatments
beyond leading logarithm in the literature for QCD hard-scattering
processes were for the independent-scattering description of hadron-hadron
elastic scattering, \refmark\BSt\ and for the scattering
of Wilson lines. \Ref\SoStKK{M.G.\ Sotiropoulos and G.\ Sterman,
 Nucl\ Phys.\ B419  (1994) 59;
B425 (1994) 489; G.P.\ Korchemsky, Phys.\ Lett.\ B325 (1994) 459;
I.A.\ Korchemskaya and G.P.\ Korchemsky, Ref.\ [\KRad].}  

In place of eq.\ \Sudconv, we now have a somewhat more general
form, in which there may be more than two jets, and in which 
the hard and soft scattering functions are labelled by their color
content,
$$
\eqalign{
{\tilde \sigma}(N) =& 
\int_0^\infty dw\, e^{-Nw}\, 
\sigma(w) \cr
=&\ \sum_{IJ}
H_{IJ}(p_1/\mu,p_2/\mu,\zeta_i)\, 
{\tilde S}_{IJ}(Q/\mu N,v_i,\zeta_i)\
\prod_i J_i(p_i\cdot \zeta_i/\mu,Q/\mu N^{1/{a_i}})\, .
}
\eqn\Sudconvcolor
$$
The indices $IJ$ refer to the color structure of the
hard scattering.  
Because soft gluons decouple from the jet functions, the jets
are diagonal in color after factorization,
just as for parton distributions.
There are
two indices in the typical cross section that describes a single
hard scattering,
one from the amplitude and one from its
complex conjugate.  
For the high-$Q^2$ elastic scattering amplitudes of protons, there
are as many as three color indices, one for each
of three quark-quark scattering amplitudes.
When the
scattering involves quarks, these color factors
may be described by singlet or octet exchange. \refmark\BSt \refmark\SoStKK
Quark-gluon and gluon-gluon scattering may be
labelled correspondingly.

For  heavy quark production, \refmark\KidSt
when the heavy quark mass is of the order of $x_1x_2s$,
with $x_i$ partonic momentum fractions, there are
only two jet factors, one for each of the incoming
partons.  The heavy quark propagators involve no
collinear divergences (and hence no Sudakov logarithms).  They
may thus be absorbed, in eikonal approximation, into the
soft tensor $S_{IJ}$.  The dynamics of the heavy quarks
beyond eikonal approximation may be absorbed into $H_{IJ}$.
For two-jet production near the elastic limit, there are four
$J_i$, two representing incoming partons, and two representing
the outgoing jets.

The analysis of Section 2 may be repeated for \Sudconvcolor,
leading to an analog of 
the resummed result \logsigmasoln, except that now,
because the anomalous dimensions corresponding to the
hard and soft functions are color matrices, we find ordered
exponentials for nonleading logarithms.
To be specific, the analogs of eqs.\ \anomdim\ for
anomalous dimensions are
$$
\eqalign{
\mu {d \over d\mu} (\ln H)_{IJ} &= 
(\Gamma_H(\alpha_s))_{IJ}\cr
\mu {d \over d\mu} \ln {\tilde J}_i &= \gamma_{J_i}(\alpha_s),\ \cr
\mu {d \over d\mu} \ln S_{IJ} &= (\Gamma_S(\alpha_s))_{IJ}\, .
}
\eqn\anomdimcolor
$$ 
The independence of $d\sigma/dw$ from
the renormalization scale $\mu$ then leads to 
$$
(\Gamma_H(\alpha_s))_{KL}+(\Gamma_S(\alpha_s))_{KL}+\sum_i \gamma_{J_i}\delta_{KL} =0\, .
\eqn\sumanomequalzerocolor
$$  
We shall not work out any example \refmark\KidSt\ of
this formalism further here.
Clearly, however, up to the matrix structure of 
the anomalous dimensions, the arguments of the previous section may 
be repeated.  In a basis in which the color anomalous
dimensions are diagonal, \refmark\BSt \refmark\SoStKK there
is a separate exponentiation for each diagonal
color structure, to leading logarithm in $N$ for $S_{IJ}$,
which is next-to-leading logarithm overall.

\chapter{Evolution in the Momentum Transfer}

The Sudakov exponentiation in eq.\ \logsigmasoln\ leads in turn to
an alternate form of resummation, from its
dependence on the momentum transfer $Q^2$.
This approach is particularly natural for electroweak processes,
in which the hard scattering is a color singlet.

We begin by taking the derivative of eq.\ \logsigmasoln\ with respect
to $\ln Q$ to derive 
the Sudakov evolution
equation
$$
\eqalign
{
Q{d\over dQ} {\tilde \sigma}(N,Q^2) 
=&
E_1(\alpha_s(\QQ))+E_2(\alpha_s(Q^2/N^2),a)
- 2\int_{Q/N^{1/a}}^Q {d\xi \over \xi}\; A(\alpha_s(\xi^2))\cr
&\quad +{2\over 1-a}\int_{Q/N^{2/a}}^{Q/N^{1/a}} {d\xi \over \xi}\; A(\alpha_s(\xi^2))\, ,
}
\eqn\sudkern
$$
where, (suppressing arguments of the functions),
$$
E_1=\beta{\partial \over \partial g}D_1+ 2B\, , \quad
E_2=\beta{\partial \over \partial g}D_2- 2B\, .
\eqn\uddef
$$
The structure of this equation is enough to imply
an alternate resummed expression, of the sort derived, for instance, in
Ref.\ [\Stdy], in which the exponent is expressed as an inverse moment.

Eq.\ \sudkern\ shows that to leading power in $N$,
any factorizing cross section obeys an evolution equation 
of the form 
$$
\QQ {d\over d\QQ} {\tilde \sigma}(N,\QQ) =
        {\tilde W}(N,\QQ) {\tilde \sigma}(N,\QQ)\, .
\eqn\twonineteen 
$$
To be specific we may take
${\sigma}(z,\QQ)$ to represent, for instance, the Drell-Yan (DY) cross section
$\sigma_{\DY}(z,\QQ)$ or a deeply
inelastic scattering (DIS) 
structure function ${\rm F}_{\DIS}(z,\QQ)$.

The inverse transform of \twonineteen\ is a convolution, whose
precise nature depends on which transform is appropriate
to the cross section.  To be specific, and to relate to the
examples we discuss in the following section, we choose
the Mellin transform, which leads to the 
familiar evolution equation\Ref\AP{G. Altarelli and G. Parisi, 
Nucl.\ Phys.\ B126 (1977) 298;
V.N. Gribov and L.N. Lipatov, Sov. J.\
Nucl.\ Phys.\ 15 (1972) 438, 675; 
 Yu.L. Dokshitser, Sov.\ Phys.\ JETP 46 (1977) 641.},
$$ 
\eqalign{
\QQ {d\over d\QQ} \sigma(z,\QQ) =&
   \int^1_z {dz'\over z'} W(z',\QQ) 
\sigma(z/ z',\QQ)\, , \cr
W(z,\QQ) =& \int_{-i\infty}^{i\infty} {dN\over 2\pi i}\, N^{-z}\;
{\tilde W}(N,Q^2)\, ,
}
\eqn\twoseventeen
$$
where in the second line we have shown the formal inverse
relation between $W$ and ${\tilde W}$.

Working in $d>4$ dimensions, 
we can take these quantities 
to be normalized such that 
$$
{\tilde \sigma}(z,\QQ=0) =
\delta(1-z)\, .
\eqn\boundry
$$ 
The physical content of this condition is that for zero
momentum transfer there is no radiation when the scattering
is through an electroweak current.  
In more than four dimensions,
these conditions are explicitly realized order-by-order in
perturbation theory. \refmark\MagSt

The solution to eq.\ \twonineteen\ 
with boundary condition ${\tilde \sigma}(N,0) =1$ is simply
$$
\eqalign{
{\tilde \sigma}(N,Q^2) 
 &\equiv{\tilde \sigma}(N,Q^2/\mu^2,\al,\ep) \cr
&= \exp\left[ \int_0^{\QQ}
 {d \xii \over \xii} 
{\tilde W} \biggl(N,{\xii\over \mu^2},\al,\ep\biggr) \right]\, ,}
\eqn\twotwenty
$$
where as above $d=4-2\epsilon>4$, and $\alpha$ is defined by
$$
\al\equiv {\alpha_s(\mu^2) \over \pi}\, .
\eqn\epsdef
$$
In eq.\ \twotwenty, we have made explicit the
perturbative dependence of ${\tilde W}$ 
on the renormalization scale $\mu$.

We can now employ the invariance of the physical quantity
 $W$ under changes in the factorization scale 
$\mu$. There
is considerable freedom in how to proceed. Since the LHS of eq.~\twotwenty\
is a renormalization group invariant, so is the exponent, and the function
$W(z,\xi^2/\mu^2,\al,\ep)$ as well.
 From eq.\ \logsigmasoln, we easily conclude
that the coefficient of $\al^M$ in ${\tilde W}(N)$ has 
 powers of $\ln N$ up to $\ln^{M+1}N$.  We can now show that
the  general 
form of the coefficient of $\alpha_s^M(\mu^2)$ in the exponent is 
$$
\eqalign{ W^{(M)}(z,\QQ/ \mu^2,\ep)&=
\sum_{j=0}^{M} c^{(M)}_j(\QQ/ \mu^2,\ep) {\cal D}_j(z) 
 + f^{(M)}(\QQ/\mu^2,\ep) \delta(1-z)  \cr
 &\quad \quad \quad \quad +
 h^{(M)}(z,\QQ/ \mu^2,\ep)\, ,
}
\eqn\twotwentyone
$$
where $h^{(M)}(z)$ is regular at $z=1$, and where
the plus distributions ${\cal D}_j(z)\equiv [\ln^j(1-z)/1-z]_+$ 
are defined as usual by
$$ 
\int_x^1 dz g(z) [f(z)]_+ = 
\int_x^1 dz (g(z)-g(1))f(z) - g(1)\int_0^x dz f(z)\, ,
\eqn\twotwentytwo
$$
with $g(z)$ an arbitrary but smooth function. 
To show \twotwentyone, we recall that for large $N$,
$$
\int_0^1 dz z^{N-1} {\cal D}_j(z) 
 = \sum_{k=0}^{j+1} t_{jk} \ln^{k}N\, , 
\eqn\twotwentythree
$$
with corrections that fall off as $1/N$,
where the matrix $[t]$ is invertible (because it is 
triangular; see for example Table 1
in\ [\CT] for explicit coefficients). Then the plus
distributions, $W_{\cal D}^{(M)}$ in eq.\ \twotwentyone\ have moments
$$
\eqalign{ 
{\tilde W}_{\cal D}^{(M)}(N,\QQ/ \mu^2,\al,\ep)
 &
= \sum_{k=1}^{M+1} \ln^{k}N \sum_{j=k-1}^M  c^{(M)}_j t_{jk} 
+\sum_{j=0}^Mc_j^{(0)}t_{j0}\cr
& \equiv \sum_{k=0}^{M+1} b_k^{(M)} \ln^k N,}  
\eqn\twotwentyfour
$$
where $b_k^{(M)} = \sum c^{(M)}_jt_{jk}$
is a function of $Q/\mu$ and $\epsilon$ in general. 

The functions 
$f^{(M)}$
and $h^{(M)}$ in eq.\ \twotwentyone\
are Mellin transforms of constants and terms behaving as
powers of $1/N$ respectively. Since the renormalization
group does not
affect the $N$ dependence, the 
$b_k$'s are RG invariant, and by invertibility of the matrix $[t]$,
so are the $c_j$'s. Then the series $f(z)=\sum_Mf^{(M)}$ and 
$h(z)=\sum_M h^{(M)}$ are also
RG invariant.
Later we will use this freedom in treating parts of the 
resummed partonic quantities differently under the
renormalization group.

For now we apply renormalization scale invariance to the full $W$
in eq.\ \twotwenty:
$$W(z,\xi^2/\mu^2,\al,\epsilon)=W(z,1/\nu,\alpha(\nu\xi^2/\mu^2,
\al,\epsilon),\epsilon)\, ,
\eqn\twotwentyfive
$$
where we are free to choose $\nu$ to be a function
of $z$ (and we shall below).
To include all corrections of the form $(\alpha\ln N)^n$ in the exponent,
we need only up to the two-loop running coupling
\refmark\CT\Ref\paperz{H. Contopanagos
 and G. Sterman, Ref.\ [\CS].}.
Here we need the dimensionally continued version of $\alpha_s$,
which we have denoted $\alpha(\lambda,\al,\epsilon)$.

The defining equation of the
$d$-dimensional running $\alpha$ is
$$
\lambda^{1-\epsilon}{\partial [\lambda^\epsilon\alpha(\lambda,\alpha(\mu^2),\epsilon)]
\over \partial \lambda}=-b_2\alpha^2(\lambda,
\alpha(\mu^2),\epsilon)-
b_3\alpha^3(\lambda,
\alpha(\mu^2),\epsilon)\, ,
\eqn\twotwentysix
$$
with boundary condition $\alpha(\lambda=1,\alpha(\mu^2),\epsilon)=\alpha(\mu^2)$.
Here $b_2 = (11 C_A - 2 n_f)/12$ and $b_3 = 34 C_A^2/48 -
(20 C_A/3 + 4 C_F) n_f/32$.
The solution, linearized in $b_3$, is:
$$ 
\lambda^\epsilon\alpha(\lambda,\alpha(\mu^2),\ep)=
{\alpha(\mu^2)\over 1-\gamma(\lambda^\epsilon,\ep)
\alpha(\mu^2)}+{b_3\over b_2}
{\alpha^2(\mu^2)\over (1-\gamma(\lambda^\epsilon,\ep)
\alpha(\mu^2))^2}f(\lambda^\epsilon,\alpha(\mu^2),\ep)\ ,
\eqn\twotwentyseven
$$
with
$\gamma(\lambda^\epsilon,\ep)\equiv {b_2\over \epsilon}(\lambda^{-\ep}-1),\ 
f(\lambda^\ep,\alpha,\ep)=1-\lambda^{-\ep}-
\left(1+{\epsilon\over b_2\alpha}\right)\ln(1-\gamma(\lambda^\ep,\ep)\alpha)$.

Given eq.\ \twotwentyfive,
the exponent in eq.\ \twotwenty\  can now be expanded as
a power series in $\alpha(\lambda,\alpha(\mu^2),\epsilon)$,
with $\lambda=\nu\xi^2/\mu^2$,
$$
\eqalign{{\tilde \sigma}(N,Q^2/\mu^2,\alpha(\mu^2),\epsilon)& =
\exp\biggl[\int_0^1dzz^{N-1}\int_0^{\nu Q^2/\mu^2}{d\lambda\over\lambda}
\{\alpha(\lambda,\alpha(\mu^2),\epsilon)W^{(1)}(z,1/\nu,\epsilon)\cr
&\hbox{\hskip 1.5 true in}  
+\alpha^2(\lambda,\alpha(\mu^2),\epsilon)W^{(2)}(z,1/\nu,\epsilon)\}
\biggr]\ .}
\eqn\twotwentynine
$$
In this resummed expression, the exponent
is itself in the form of a moment.\refmark\Stdy  A similar analysis
may be given for Fourier transforms. \Ref\AGM{G.\ Altarelli, R.K.\ Ellis, 
M.\ Greco and G.\ Martinelli, Ref.\ 
[\qtresum].}\Ref\KS{G.P. Korchemsky and G. Sterman,
Ref.\ [\IRR].} 
The $d$-dimensional running coupling $\alpha$ in eq.\ \twotwentynine\ 
has a usual perturbative (``renormalon"\refmark\IRR)
singularity at $\lambda=\exp[-1/\alpha b_2]$, as seen from
eq.\ \twotwentyseven.  Eq.\ \twotwentynine\ must therefore be considered as a 
resummation order-by-order in a perturbative expansion of
its exponent, without further information on
how to treat the running coupling at very low scales.

The functions $W^{(1)},\ W^{(2)}$ 
can be determined by choosing $\nu=1$ in eq.~\twotwentynine,
and expanding it to two loops in $\al$,
$$
\eqalign{
Q^2{\partial \over \partial Q^2}\; 
\ln\; {\tilde \sigma}(N,Q^2/\mu^2,\alpha(\mu^2),\epsilon)
= &
\alpha(Q^2/\mu^2,\alpha(\mu^2),\epsilon){\tilde W}^{(1)}(N,1,\epsilon) \cr
& \quad\quad + \alpha^2(Q^2/\mu^2,\alpha(\mu^2),\epsilon)
{\tilde W}^{(2)}(N,1,\epsilon)\, ,
}
\eqn\twothirty
$$
in terms of $\alpha_s(\mu^2)$, using (from \twotwentyseven) 
$$
\lambda^{\epsilon}\alpha(\lambda,\alpha(\mu^2),\ep)=
\alpha(\mu^2)
+\gamma(\lambda^\epsilon,\ep)\alpha^2(\mu^2)\ .
\eqn\twothirtyone$$
We then find for the perturbative expansion of the moments
$$
{\tilde W}^{(1)}(N,1,\ep)=
(Q^2/\mu^2)^{\epsilon}\, 
Q^2 {\partial \over \partial Q^2}\, 
{\tilde \sigma}^{(1)}(N,Q^2/\mu^2,\epsilon)\ ,
\eqn\twothirtyonehalf
$$
and 
$$
\eqalign{
{\tilde W}^{(2)}(N,1,\ep) =& 
(Q^2/\mu^2)^{2\epsilon}\, \Bigg \{
{\partial \over \partial \ln Q^2}\, \biggl (
{\tilde \sigma}^{(2)}(N,Q^2/\mu^2,\epsilon) 
-{1\over 2} \biggl [{\tilde \sigma}^{(1)}(N,Q^2/\mu^2,\epsilon)\biggr ]{^2}
\biggr ) \cr
& \hbox{\hskip 1.0 true in}
- \gamma((Q^2/\mu^2)^\epsilon,\epsilon){\partial \over \partial \ln Q^2}
{\tilde \sigma}^{(1)}(N,Q^2/\mu^2,\epsilon) \bigg \}\, .
}
\eqn\twothirtytwo
$$
These results may be inverted to derive $W^{(1,2)}(z,1,\epsilon)$.  
The full functions $W^{(1,2)}(z,\xi^2/\mu^2,\epsilon)$ may then
be constructed by reexpanding $\alpha(\xi^2/\mu^2,\alpha(\mu^2),\ep)$
in $\alpha(\mu^2)$ in eq.\ \twotwentyfive, using \twotwentyseven\ with 
$\lambda=\xi^2/\mu^2$,
$$
\eqalign{
W^{(1)}(z,\xi^2/\mu^2,\epsilon)
&=
\bigg ({\mu^2 \over \xi^2}\bigg)^\epsilon W^{(1)}(z,1,\epsilon)\, ,
\cr
W^{(2)}(z,\xi^2/\mu^2,\epsilon)&= 
\bigg ({\mu^2 \over \xi^2}\bigg)^{2\epsilon} W^{(2)}(z,1,\epsilon)
+\gamma((\xi^2/\mu^2)^\epsilon,\epsilon)\,
\bigg ({\mu^2 \over \xi^2}\bigg)^\epsilon W^{(1)}(z,1,\epsilon)\, .
}
\eqn\wcuefromwone
$$
According
to eqs.~\twotwentynine\ and \twothirtyonehalf\ 
the complete one loop result exponentiates
in moment space \refmark\Stdy,
up to possible corrections that behave as  $1/N$.
By comparison of 
eq.\ \twotwentynine\ with the general 
resummed expression in moment space, eq.\ \logsigmasoln,
we observe that we are able to choose $\nu$ in
such a way as to absorb all logarithmic $1-z$
dependence into the running coupling.  When this is done,
leading and next-to-leading logarithms in $N$ 
will be resummed by the $W^{(1,2)}$ terms in
\wcuefromwone.  We shall make use of this 
result in the following section.

\chapter{Example:  Resummation for Inclusive DIS and DY}

As a final topic, we illustrate the simplicity of the
resummation method of the previous section, by applying
it directly to the Drell-Yan and deeply inelastic scattering
cross sections.  We go on to verify that the resulting
expressions reproduce known finite-order and resummed
expressions for the hard-scattering function in the
Drell-Yan process in DIS and $\overline{\rm MS}$ schemes.

Consider first inclusive DIS,
$$V(q) + h(P)\rightarrow X, \eqn\twoeleven$$
where $V$ is a vector boson with spacelike momentum $q$,
and $h$ is a hadron. The cross section for this
process is customarily expressed in terms of structure
functions $F_i$ ($i=1,2,3$). We focus on the case 
$V = \gamma^*$, and the function $F_2$.
It can be factorized in moment space, as noted in the introduction,
$$
F_2^{(h)}(N,Q^2)= \int_0^1 dx \,x^{N-1} 
F_2(x,\QQ)=\sum_i C_i(N,Q^2)\, \phi_{i/h}(N,Q^2)\, ,
\eqn\momentDIS
$$
where $\QQ = -q^2$ and $x = \QQ/2 P\cdot q$. Here the $C_i$
are partonic coefficient functions, and the $\phi_{i/h}$ parton
density functions for hadron $h$. 
(In the following, we denote Mellin transforms by their
arguments, and drop the tilde of previous sections.)
We will examine $F_2^{(h)}$ for large 
$N$, where the quark contribution $i=q$ 
is dominant. Henceforth
we will denote the contribution 
of a single quark to $F_2^{(h)}$ by ${\rm F}_{\DIS}$, dropping
the hadron label.
By itself, \momentDIS\ does not yet lead to
Sudakov exponentiation in moment space.  In particular
$C$ is still a function of $N$.  
${\rm F}_{\DIS}$, however, obeys another factorization theorem
near the edge of phase space, i.e. at large $N$ 
(eq. (3.14) in Ref.\ [\Stdy]),
$$
{\rm F}_{\DIS}(N,\QQ) =\bigg |{\rm H}_{\DIS}\biggl({\QQ\over \mu^2}\biggr)
\bigg|^2 
\phi'\biggl({p_1\cdot \zeta_1 \over \mu},{Q\over \mu N}\biggr)
J\biggl({p_2\cdot \zeta_2 \over \mu},{Q\over \mu N^{1/2}}\biggr) 
V \biggl ({Q\over \mu N}\biggr ) \;\; + \;\; {\cal O}({1/N})\, , 
\eqn\twothirteen
$$
in terms of a modified quark distribution $\phi'$.
$p_1$ is the momentum of the incoming hadron (parton,
in perturbation theory), and $p_2$ the
momentum of the scattered quark in the elastic limit.
In \twothirteen,
the $N$-dependence of the 
partonic coefficient function $C_q$ and the distribution
$\phi_{q/h}$ has been factorized into $\phi'$,
a hard scattering function $H$, a jet function $J$, corresponding
to the scattered quark, and a soft function $V$.  Each of these
functions is infrared safe, but $\phi'$, $J$ and $V$ absorb all of the 
$N$-dependence, and hence singularities in the elastic
limit.  
The respective $N$ dependences of $\phi'$, $J$ and $V$ may be found
from eqs.\ (4.7), (6.1) and (7.9) of [\Stdy].
As above, the $\zeta_i$ may be thought of as gauge-fixing
vectors used to define the jet functions.
The factorized expression \twothirteen\ is of the form of eq.\ 
\Sudconv, and following the arguments of Sections 2 and 4, 
this alone is enough to ensure that
${\rm F}_{\DIS}$ enjoys a Sudakov resummation in logarithms of $N$.

Analogous arguments hold for the Drell-Yan process,
$$ 
h_A(p_1) + h_B(p_2) \rightarrow V(q) + X\, ,
\eqn\twoone
$$
where $h_A$ and $h_B$ are hadrons 
and $V$ is a heavy vector boson ($\gamma^*,Z,...$) of
timelike momentum $q$.
The factorization theorem for this process \refmark\CSSrv reads in moment space
$$ 
\eqalign{
\int_0^1 d\tau \tau^{N-1} {d\sigma \over d\QQ}(\tau, \QQ) 
 &=   \sum_{i,j}\phi_{i/A}(N,\QQ) \, \phi_{j/B}(N,\QQ) \sigma_{ij}(N, \QQ) \cr
&= \sum_q \bigg |{\rm H}_{\DIS}\biggl({\QQ\over \mu^2}\biggr)
\bigg|^2 
\psi_{q/A}\biggl ( p_1\cdot \zeta_1,{Q\over \mu N} \biggr )
U \biggl ( {Q\over \mu N} \biggr )
\psi_{{\bar q}/B}\biggl ( p_2\cdot \zeta_2,{Q\over \mu N} \biggr )
\cr
& \hbox{\hskip 1.5 true in}+{\cal O}(1/N)\, ,
}
\eqn\twothree
$$
where $\QQ=q^2$, $\tau \equiv \QQ/S$ and $S=(p_1+p_2)^2$,
and the sum is over quarks and antiquarks in the second line.
As above, the $\phi_{i/A}(x,\QQ)$ in the first 
equality are standard parton densities, and
$\sigma_{ij}$ is the corresponding partonic DY coefficient function.
Furthermore, 
$\sigma_{ij}(N,\QQ) \equiv \int_0^1 dz z^{N-1} \sigma_{ij}(z, \QQ)$,
and similarly for the $\phi$'s.
We restrict ourselves to $ij = q\bar{q}$,
which is the dominant production channel 
at large $N$, and denote 
the partonic coefficient function for this case by $\sigma_{\DY}$.
As for $F_{\rm DIS}$, $\sigma_{\DY}(z)$ is singular when $z\rightarrow 1$,
but satisfies a further factorization \refmark\Stdy, 
given in the second line of \twothree, in which all
of the $\ln N$-dependence of its moment is absorbed into 
jet functions $\psi$
 and a soft function $U$, precisely as in eq.\ \Sudconv\ above.  
The explicit $N$-dependeces
of $\psi$ and $U$ may be found from eqs.\ (5.7)
and (7.8) of [\Stdy].

Finally, the $\overline{\rm MS}$ distribution \Ref\Baretal{W.A.\ Bardeen,
A.J.\ Buras, D.W.\ Duke and T.\ Muta, Phys.\ Rev.\ D18 (1978) 3998.}, which
we denote $\phi_{\MSb}$,
may also be treated in 
a similar fashion.  This is because it differs from the  ${\rm F}_{\DIS}$ 
as  $x\rightarrow 1$ only by replacing the scattered quark jet by an
outgoing Wilson line \Ref\CollinsSoperPDF{J.C.\ Collins and 
D.E.\ Soper, Nucl.\ Phys.\ B194 (1982) 445.}.  We shall observe
below, however, certain differences in the resummation of this
function, compared to cross sections.

We are now ready to derive specific resummed expressions
for the DY and DIS processes, as well as for 
the $\overline{\rm MS}$ distribution. 
We will also determine 
the exponentiating two-loop coefficients
in the hard-scattering function of DY in the 
DIS and $\overline{\rm MS}$ factorization schemes.

According to the discussion of Section 4
we need the one loop corrections to 
$\sigma_{\DY}$, ${\rm F}_{\DIS}$ and ${\phi}_{\MSb}$.
They are in 
$d=4-2\epsilon$ ($\epsilon < 0$) dimensions 
\Ref\AEM{G.\ Altarelli, R.K.\  Ellis and G.\ Martinelli, Nucl.\ Phys.\ 
B143
(1978) 521, (E)146 (1978) 544, B157 (1979) 461.}
$$\eqalign{
  \sigma_{\DY}^{(1)}(z,\QQ)
&  = - {\alpha_s C_F\over \pi} \biggl( {4\pi\mu^2 \over \QQ} \biggr)^\epsilon
  {\Gamma(1-\epsilon)\over \Gamma(1-2\epsilon)} \Biggl[
   \delta(1-z) \biggl({1\over 2}-{\pi^2\over 3}\biggr)  \cr
& \quad\quad +{1\over \epsilon}\biggl({1 + z^2 \;\;\> \over (1-z)^{1 + 2\epsilon}}
  \biggr)_+ z^\epsilon  \Biggr] , \cr  
{\rm F}_{\DIS}^{(1)}(z,Q^2/\mu^2,\ep) 
&={\alpha_s C_F\over 2\pi}\biggl({4\pi \mu^2\over Q^2}\biggr)^\ep
 {\Gamma(1-\epsilon)\over \Gamma(1-2\epsilon)}
\Biggl[\delta(1-z)\biggl(-{11\over 4} -{\pi^2\over 3} \biggr)\cr
& \quad \quad -{1\over \ep}\biggl({1+z^2+3\ep/2\over (1-z)^{1+\ep}}\biggr)_+
z^\ep +z^\ep(1-z)^{-\ep}(3-z)+3z\Biggr]\ ,
}
\eqn\threetwo$$
and 
$${\phi}_{\MSb}^{(1)}(z,\QQ)
   = -{\alpha_s C_F\over 2\pi} \biggl( {4\pi\mu^2 \over \QQ} \biggr)^\epsilon
  {\Gamma(1-\epsilon)\over \Gamma(1-2\epsilon)}
       {1\over \epsilon} \Biggl[ {1 + z^2  \over 1-z}
   \Biggr]_+ . \eqn\threefive$$
Here $Q^2$ denotes the heavy vector boson mass squared in the
case of DY, and $-Q^2$ the invariant mass of the 
photon probe in the DIS (and $\msb$) case;
$\mm$ is the dimensional regularization scale, while
$z$ denotes the square of the ratio of the heavy vector
boson mass to the parton center-of-mass energy in DY, and the
partonic Bjorken scaling variable for DIS and  $\overline {\rm MS}$.
The plus distributions in the above expressions are usually expanded as
$$  
\Biggl[{1 \; \> \over (1-z)^{1+\kappa}}\Biggr]_+ 
 = \Biggl[{1 \; \> \over (1-z)}\Biggr]_+  -
 \kappa \Biggl[{\ln(1-z)  \> \over (1-z)}\Biggr]_+  + 
 {1\over 2} \kappa^2 \Biggl[{\ln^2(1-z) \> \over (1-z)}\Biggr]_+ 
  - \quad \dots \quad ,
\eqn\threeseven
$$
which we choose not to do here. Redefining
$\mu^2 \rightarrow \mu^2\exp (-(\ln(4\pi) - \gamma_E))$
eliminates the $(4\pi)^{\epsilon}$ in the
expressions \threetwo\ , and
absorbs the factors $\Gamma(1-\eps)/\Gamma(1-2\eps)$.

Let us exhibit the case of DIS in more detail, as an example.
The one loop correction to the DIS partonic structure function
leads via eq.~\twothirtyonehalf\
to $W_{\DIS}^{(1)}(z,1,\ep)$ in momentum space. We can write this
kernel as
$$
W_{\DIS}^{(1)}(z,1,\ep)=\delta(1-z)f^{(1)}_{\DIS}(\ep)
+z^\ep\left({g^{(1)}_{\DIS}(z,\ep)\over (1-z)^{1+\ep}}\right)_+
+h^{(1)}_{\DIS}(z,\ep)\ ,
\eqn\threeten
$$
where the coefficient functions $f^{(1)}_{\DIS},\ g^{(1)}_{\DIS},
\ h^{(1)}_{\DIS}$
are regular functions of their arguments at $z=1$, given by
$$f^{(1)}_{\DIS}(\ep)={C_F\over 2}\ep\biggl({11\over 4}+
{\pi^2\over 3}\biggr)\ ,
\eqn\threeeleven$$
$$g^{(1)}_{\DIS}(z,\ep)=
{C_F\over 2}\biggl(1+z^2+{3\ep\over 2}+{7\ep^2\over 2}\biggr)\, ,
\eqn\threetwelve$$
$$h^{(1)}_{\DIS}(z,\ep)=-{C_F\over 2}\ep[z^\ep(1-z)^{-\ep}(3-z)+3z]\ .
\eqn\threethirteen
$$
We can now use \wcuefromwone\ to determine the one-loop exponent, and
  substitute the result into the exponential
resummed cross section \twotwentynine.

Given the limitations of 
the factorization theorem \twothirteen, we 
can discard terms of order $1/N$. 
Thus we drop $h^{(1)}_{\DIS}(z,\ep)$
completely and expand $g^{(1)}_{\DIS}(z,\ep)$ around $z=1$,
dropping $1/N$ terms in this expansion where convenient. 

According to the discussion
below \twotwentythree,
 the term $\delta(1-z)\; f^{(1)}_{\DIS}$ and the plus distributions
in \threeten\ are 
 separately RG invariant.  
We are therefore free to choose different scales $\nu$ for these
two terms 
in the general resummed expression eq.\ \twotwentynine.
The natural choice for the $\delta(1-z)\; f^{(1)}_{\DIS}$ term
is  $\nu=1$ in \twotwentynine. Changes in $\mu$ generate terms
$b_2\ln(\mu'/\mu)$ at higher orders.
The $\lambda$ integral may then be carried out explicitly
for this term in \twotwentynine.
For the plus distribution term, however, the natural
choice is $\nu=1-z$.  Then,
using $\xi=(1-z)^{1/2}\mu$ in eq.\ \wcuefromwone, we absorb the factor
$(1-z)^{-\epsilon}$ in \threeten\ into the 
limit of the $\lambda$ integral in eq.\ \twotwentynine, involving
only the
running coupling, as indicated at the end of
Sec.\ 4.


Treating the $N$-independent
term $f^{(1)}$ in this fashion, we obtain for the
one-loop resummation of the DIS cross section,
$$
\eqalign{&{\rm F}_{\DIS}(N,Q^2/\mu^2,\al,\ep) 
\cr
&\quad = \exp\Biggl[{C_F\over 2}
 \int_0^1 dz 
 {z^{N-1+\ep}-1 \over 1-z}
\Biggl\{\int_0^{\QQ (1-z)/\mu^2} {d\lambda \over \lambda}
\; \biggl [ (1+z^2)+{3\epsilon \over 2} \biggr ]
\; \alpha(\lambda,\al,\ep)\, \Biggr\}\Biggr] \cr
 & \quad\quad \times \exp\left[-{\al C_F \over 2 }
(11/ 4 + \pi^2 / 3)\right]  \cr
&\quad = \exp\Biggl[{C_F\over 2}
 \int_0^1 dz 
 {z^{N-1+\ep}-1 \over 1-z}
(1+z^2)\Biggl\{\int_0^{\QQ (1-z)/\mu^2} {d\lambda \over \lambda}\; 
 \alpha(\lambda,\al,\ep) \cr
&\quad\quad\quad -{3\over 2}
\alpha\bigl(Q^2(1-z)/\mu^2,\alpha_s(\mu^2),\epsilon \bigr )\, 
\Biggr\}\Biggr] \cr
 & \quad\quad \times \exp\left[-{\al C_F \over 2 }
(11/ 4 + \pi^2 / 3)\right]\, ,
}
\eqn\threenineteen 
$$
where in the second form we have evaluated the $3\epsilon/2$
term to lowest order in $\alpha_s(Q^2(1-z))$.
Note that as long as were are in $d$ dimensions the $\lambda$ integrals
in the exponent are well-defined at their lower limits.
There is no particular problem with integrals dominated by
their 
lower limits here, since $F_{\DIS}$ is not by itself infrared
safe.

For the Drell-Yan cross section one obtains by essentially identical methods,
$$\eqalign{&\sigma_{\DY}(N,\QQ/\mu^2,\al,\ep) \cr
&\quad  = \exp \left[ C_F 
 \int_0^1 dz\;
 \left ( {z^{N-1+\ep}-1 \over 1-z}\right )\; (1+z^2)\;
 \int_0^{\QQ (1-z)^2/\mu^2} {d\lambda \over \lambda} \;
\alpha(\lambda,\al,\ep)
\right]   \cr
 & \quad \quad \times \exp\left[-\al C_F 
(1/ 2 - \pi^2/ 3)\right] \ .}
\eqn\threetwentysix$$
Here, the natural choice of scale for the 
plus distributions is $\nu=(1-z)^2$.

It is not possible to treat the $\overline{\rm MS}$ quark distribution
$\phi_{\MSb}$ in quite the same way, because it is not normalized at $Q^2=0$
by electroweak current conservation.  
Indeed, it is usually regarded as a function of only a single
scale, through the running coupling.
Nevertheless, it satisfies an evolution equation
\refmark\AP\ of precisely the form \twonineteen, with $W_{\MSb}$ given by 
the standard splitting functions.
\refmark\CollinsSoperPDF\  The problem we must therefore
solve is how to fix a boundary condition for this equation.
To this end, we note that for $\epsilon<0$ ($d>4$), the 
dimensionally-continued running coupling vanishes at zero scale,
$\alpha(0,\alpha(\mu_1^2),\eps)=0$, order by order in its perturbative
expansion in the coupling $\alpha(\mu_1^2)$ evaluated at any
nonzero scale $\mu_1$.  Dimensionally-continued radiative
corrections to $\phi_{\MSb}(N,\alpha(Q^2))$ therefore vanish at $Q^2=0$,
and we may take, just as for the electroweak cross sections above,
$\phi_{\MSb}(N,Q^2=0,\eps)=0$.  The solution to the evolution equation
is  then fixed by this boundary condition, and we find
$$\eqalign{
& {\phi}_{\MSb}
(N,Q^2,\epsilon) \cr
& \quad = \exp\left[ -\int_0^{Q^2}{d\mu'{^2}\over \mu'{^2}}\, \Gamma_{qq}(N,\alpha_s(\mu'{^2}))
\;\right ]
\cr
& \quad =\exp\left[{C_F\over 2}
\int_0^1dz\left({z^{N-1}-1\over 1-z}\right)\; (1+z^2)\;
\int_0^{Q^2/\mu^2}{d\lambda\over \lambda} \;\alpha(\lambda,\alpha
(\mu^2),\epsilon) +\dots\; \right]\ .}
\eqn\threetwentyone$$
In the first equality, 
 $\Gamma_{qq}(N,\alpha)$ are the (singular, diagonal) terms in the
quark anomalous dimension matrix, which contribute $\ln N$ dependence.
In the second equality, we have introduced an arbitrary scale $\mu^2$
through $\lambda\equiv \mu'{^2}/\mu^2$, for which $\alpha(\lambda,\alpha(\mu^2),\eps)
= \alpha(\mu'{^2})$.  This brings our
expression into the same form as eqs. \threenineteen\ and \threetwentysix.
   
The expressions for ${\rm F}_{\DIS}$, ${\phi}_{\MSb}$ and
$\sigma_{\DY}$ in eqs. \threenineteen, \threetwentyone\ and \threetwentysix\ 
are singular as $\ep\rightarrow 0$. 
However, this singular behavior is universal for all three 
quantities. Thus we can perform mass
factorization to obtain the finite Drell-Yan hard part,
which is a simple ratio in moment space:
$$
\omega_{\qqb}(N,\QQ/\mu^2,\al) = 
{\sigma_{\DY}(N,\QQ/\mu^2,\al,\ep) \over
 \phi^2(N,\QQ/\mu^2,\al,\ep)}\, ,
\eqn\threetwentyseven
$$
where $\phi(N)$ is the moment of  a suitable quark distribution
(mixing with other partons is down by $1/N$).  
Choosing for $\phi$ alternatively ${\rm F}_{\DIS}$ and
${\phi}_{\MSb}$ in \threetwentyseven\  yields the finite resummed hard part
for the Drell-Yan cross section in the
$\dis$ and $\msb$ schemes. The factorization scale is set
by the denominator, and we choose it to be $Q^2$. 
Putting $\mu^2=Q^2$, and returning to four dimensions
because $\omega$ is infrared safe, we have finally
$$
\eqalign{
        \omega^{\DIS}_{\qqb}(N,\alpha(Q^2)) =&
\exp\Biggl[ \int_0^1 dz \; \left ( {z^{N-1}-1\over 1-z}\right )\;
\Biggl\{ (1+z^2)\; \int_{(1-z)^2}^{(1-z)}
        {d\lambda \over \lambda} 
        (-2C_F)\alpha(\lambda Q^2) \cr
& \quad\quad +  {3 C_F\over 2} 
        \alpha((1-z)Q^2)\Biggr\} \Biggr] \cr
&\times  
\exp\left[  {\alpha(Q^2) C_F \over 2}(9/2 
+ 4\pi^2/3)\right] \ ,}
\eqn\threetwentyeight
$$
and
$$\eqalign{
        \omega^{\MSb}_{\qqb}(N,\alpha(Q^2))  =& 
\exp\left[  \int_0^1 dz\; \left ( {z^{N-1}-1\over 1-z} \right )\;
(1+z^2)\; \int_{(1-z)^2}^1 
        {d\lambda \over \lambda} 
        (-2C_F)\alpha(\lambda Q^2) 
        \right]   \cr
\times & 
\exp\left[  {-\alpha(Q^2) C_F \over 2}(1 - 2\pi^2/3)\right]\ .}
\eqn\threetwentynine$$
One may check that expansion to one loop reproduces 
the one loop hard parts in [\AEM],
up to terms of order $1/N$. We denote
the $N$-independent terms in the above expressions
 by $A(\alpha(\QQ)). $\foot{Note that they may
be numerically quite important.}
Then both of the above results are of the form 
$$
\omega_{\qqb}^S(N,\alpha(Q^2))=A_S(\alpha(Q^2))
I_S(N,\alpha(Q^2))\ ,
\eqn\threethirty
$$
where the $N$-dependent exponential $I_S$ is
$$
\eqalign{
&I_S(N,\alpha(Q^2))=\cr
&\exp\left[-\int_0^1dz{z^{N-1}-1\over 1-z}\Biggl\{
\int_{(1-z)^2}^{(1-z)^{m_S}}{d\lambda\over \lambda}g_1[\alpha(\lambda Q^2)]
+g_2[\alpha((1-z)^{m_S}Q^2)]\Biggr\}\right]\, .}
\eqn\threethirtyone
$$
Here $m_S=1,0$ for the  $\dis$ and $\msb$ schemes respectively,
and the functions $g_1,\ g_2$ 
have finite expansions in their arguments:
$$
g_1[\alpha]=\sum_{n=1}^\infty\alpha^ng_1^{(n)}\ ,\ g_2[\alpha]=
\sum_{n=1}^\infty\alpha^ng_2^{(n)}\ .
\eqn\threethirtytwo
$$
The expression \threethirtyone\ organizes all
large logarithms ($\ln N$ or $\ln(1-z)$) in the perturbation
expansion for $\omega_{\qqb}^S$. Its expansion in $\alpha(Q^2)$ generates
terms $\alpha^n(Q^2) \ln^mN$, $n=0,..,\infty, \; m \leq n+1$
in the exponent. 
To be specific, for a given order $n$, one can call terms for which
$m= n+1$ ``leading logarithmic", terms with
$m=n$ ``next-to-leading", and ones where
$m<n$ ``subdominant". In this terminology $g_1^{(1)}$
contributes at the leading logarithmic level, 
$g_2^{(1)}$ and $g_1^{(2)}$ at the next-to-leading level, while 
$g_2^{(2)}$ is subdominant.

The form \threethirtyone\ was derived via different methods 
in refs.\ [\Stdy] and [\CT]\ for the $\dis$ scheme, and 
for the $\msb$ scheme in Ref.\ \REF\Catmsb{S. Catani, G. Marchesini,
B.R. Webber, Nucl.\ Phys.\ B349 (1991) 635.}\ [\Catmsb].
The lowest order coefficients are obvious from \threetwentyeight\ 
and \threetwentynine:
$$g_1^{(1)}|_{\DIS}=2C_F\ ,\ g_2^{(1)}|_{\DIS}=-3C_F/2\eqn\threethirtythree$$
$$g_1^{(1)}|_{\MSb}=2C_F\ ,\ g_2^{(1)}|_{\MSb}=0\ ,\eqn\threethirtyfour$$
and agree with those found in [\Stdy], [\CT]\ and [\Catmsb].
Although, strictly speaking, we have not proved here that the form
\threethirty\ holds for all $g_i^{(j)}$,
functions of different arguments for the running coupling
can always be brought to the above form, by 
the method discussed in Ref.\ [\Catcomm].  In particular, the
argument of $\alpha_s$ in $g_2$ in eq.\ \threethirtyone\ is arbitrary in
the ${\overline{\rm MS}}$ scheme, up to subdominant terms,
because $g_2^{(1)}|_{\MSb}=0$.

We may improve the accuracy of our resummation by
determining the coefficients $g_i^{(2)}$ in both
schemes. 
We do this by expanding \threethirtyone\ to order $\alpha^2$ and comparing
with the explicit two-loop results in
\REF\Neerven{T. Matsuura, S.C. van der Marck and W.L. van Neerven,
Nucl.\ Phys.\ B319 (1989) 570;
R. Hamberg, W.L. van Neerven and T. Matsuura, Nucl.\ Phys.\ B359 (1991)
343.}\ [\Neerven],
which is most conveniently done in moment space.
Using methods described in \REF\Magnea{L. Magnea, Nucl.\ 
Phys.\ B349 (1991) 703.}\ [\Magnea], we find,
as in [\CT], [\Magnea],
$$g_1^{(2)}|_{\DIS}=g_1^{(2)}|_{\MSb}=\gamma_K^{(2)}
=C_A C_F\biggl({67\over 18}-\zeta(2)\biggr) + n_f C_F \biggl(-{5\over 9}\biggr)
, \eqn\threethirtysix$$
the two loop anomalous dimension first identified by
Kodaira and Trentadue \Ref\KT{J. Kodaira and L. Trentadue, 
Phys.\ Lett.\ B112 (1982) 66.}
\Ref\DaviesStir{C.T.H. Davies and W.J. Stirling, in Ref.\ [\qtresum].}
\refmark\Collinsrv\refmark\KRad, and
$$
\eqalign{& g_2^{(2)}|_{\DIS} = C_F^2\left( -{3\over 16} + {3\over 2}
\zeta(2) - 3\zeta(3)\right) + C_A C_F \left(-{57\over 16} -{11\over 6}\zeta(2)
+ {3\over 2}\zeta(3)\right)  \cr
& \quad\quad+ n_f C_F \left({5\over 8} + {1\over 3}\zeta(2)\right) \cr
& g_2^{(2)}|_{\MSb} = C_A C_F \left({101\over 27} -{11\over 3}\zeta(2)
- {7\over 2}\zeta(3)\right) + n_f C_F \left(-{14\over 27} 
+ {2\over 3}\zeta(2)\right). }
\eqn\threethirtyseven
$$
The numerical impact of the latter  coefficients is quite small
near threshold,
as they contribute only at the sub-dominant logarithmic level.


\chapter{Conclusion}

We have presented a unified method for resumming Sudakov corrections to
 partonic cross sections. We assumed the existence of a factorized
expression \Sudconv, which is applicable to a large class of cross sections
in the ``elastic limit" of phase space, where their
QCD corrections due to soft gluons can be large.
To factorize the cross section in these regions, it is in general
necessary to introduce
dependence on auxiliary vectors $\zeta_i$. 
The freedom in choosing the $\zeta_i$ leads to 
consistency equations which resum
the large corrections. 

To illustrate the method we have rederived the large $x$
resummed Drell-Yan hard parts in $\dis$ and ${\overline{\rm MS}}$ factorization
schemes.  These expressions are relevant to
both perturbative resummation and power corrections. For example, in \ [\CS],
the $z$ integral in \threethirty\  was treated
 by a principal value presciption \Ref\Mueller{A.H.\ Mueller,
Nucl.\ Phys.\ B250 (1985) 327.}
for the Drell-Yan process in DIS scheme, to define the integral
in the presence of its 
renormalon pole \refmark\IRR.  It was observed that in resummed perturbation
theory the pole manifests itself only
at the level of power-suppressed (``higher twist") effects. 
The precise nature of this dependence is an active subject of inquiry.
Comparisons with data using the
resummed (differential) DY cross section have also been
carried out.
\refmark\AC\
Other questions of theoretical interest may include the
application of effective field theory methods to Sudakov factorization.
Phenomenologically interesting applications
are possible to any QCD reaction for which
soft gluon resummation is relevant,
such as high $p_t$ jet production and high mass heavy
quark production. \refmark\QQprod\refmark\KidSt
We hope that the streamlined discussion
of Sudakov resummation given above will facilitate future
inquiry in this area.

\bigskip

\noindent
{\it Acknowledgements.}  We would like to thank
Lyndon Alvero,
Ed Berger, John Collins, Nicholaos Kidonakis,
Gregory Korchemsky, Al Mueller, 
Jack Smith, Dave Soper
and Wu-Ki Tung for helpful conversations.
This work was supported in part 
by the National Science Foundation under
grant PHY9309888, and by the U.S. Department of Energy under contract
W-31-109-ENG-38.

\refout
\endpage
\title {{\bf FIGURE CAPTION}} 

Figure 1. (a) Cut diagram representing the factorization of
eq.\ \Sudconv\ in ${\rm e}^+{\rm e}^-$ annihilation in the
elastic limit;  (b) same for DIS; (c) same for Drell-Yan.

\endpage

\vglue 12cm
\vbox{\includegraphics{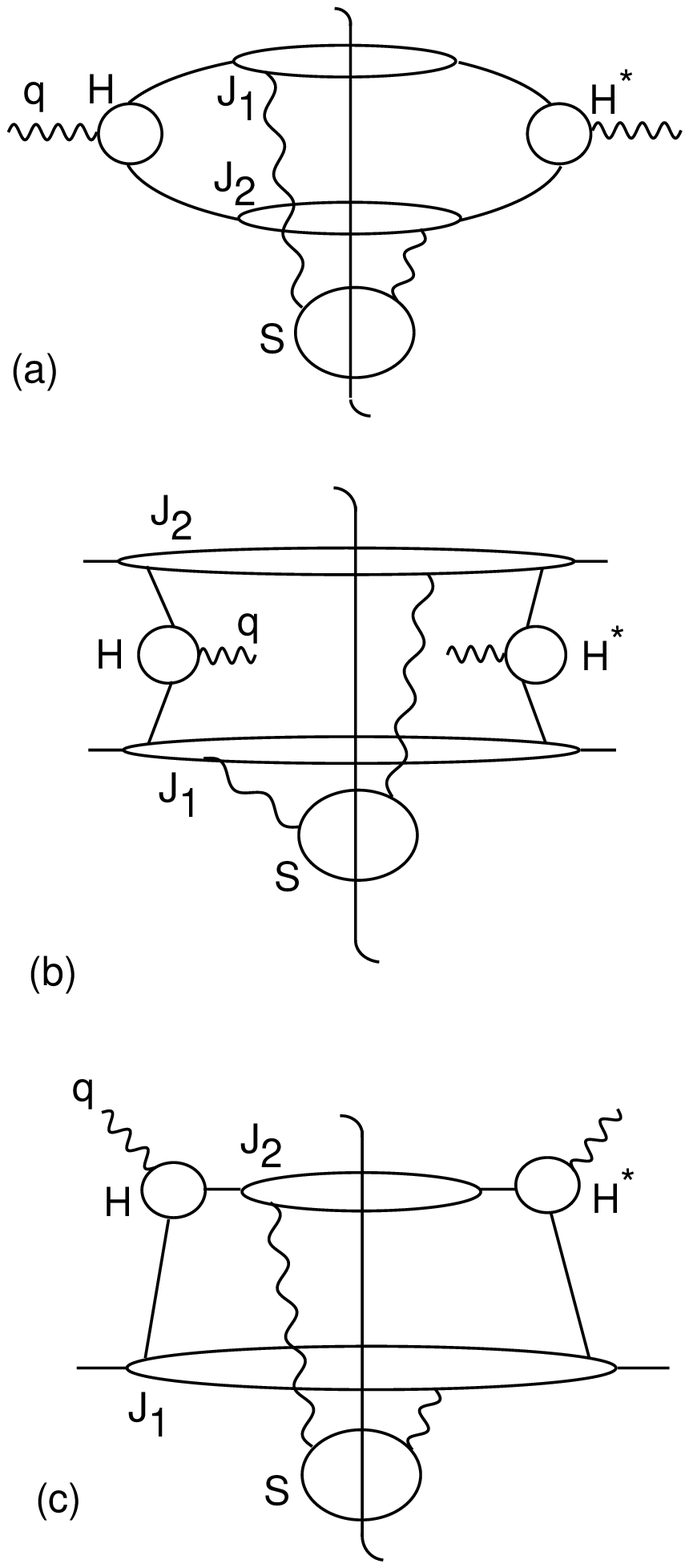} }

\nopagenumbers

\end
\bye